\documentclass[prl,twocolumn,superscriptaddress,nobibnotes,a4paper]{revtex4-2}
\usepackage[pdftex]{graphicx}
\usepackage[pdftex]{epsfig}
\usepackage{amsmath}
\usepackage{amssymb}
\usepackage{amsfonts}
\usepackage{color}
\usepackage{wrapfig}
\usepackage{eucal}
\usepackage{hhline}
\usepackage{threeparttable}
\usepackage{supertabular}
\usepackage{multirow}
\usepackage{tabularx}
\usepackage{siunitx}
\usepackage{upgreek}
\usepackage{float}

\usepackage[export]{adjustbox}
\usepackage{xcolor}
\usepackage{soul}

\usepackage[version=4]{mhchem} 
\usepackage{array} 
\usepackage[colorlinks=true, allcolors=blue]{hyperref} 

\newcommand{\und}[1]{_\textrm{#1}}

\setstcolor{red}

\begin{document}

\title{Broadband, High-Reflectivity Dielectric Mirrors at Wafer Scale: Combining Photonic Crystal and Metasurface Architectures for Advanced Lightsails}

\author{Jin Chang}
\affiliation{Kavli Institute of Nanoscience, Department of Quantum Nanoscience, Delft University of Technology, 2628CJ Delft, The Netherlands}
\author{Wenye Ji}
\affiliation{Department of Imaging Physics, Delft University of Technology, Lorentzweg 1, 2628CJ, Delft, The Netherlands}
\author{Xiong Yao}
\affiliation{Kavli Institute of Nanoscience, Department of Quantum Nanoscience, Delft University of Technology, 2628CJ Delft, The Netherlands}
\affiliation{Faculty of Physics, School of Science, Westlake University, Hangzhou 310030, P.R. China}
\affiliation{Department of Physics, Fudan University, Shanghai 200438, P.R. China}
\author{Arnold J. van Run}
\affiliation{Kavli Institute of Nanoscience, Delft University of Technology, 2628CD, Delft, The Netherlands}
\author{Simon Gr\"oblacher}
\email{s.groeblacher@tudelft.nl}
\affiliation{Kavli Institute of Nanoscience, Department of Quantum Nanoscience, Delft University of Technology, 2628CJ Delft, The Netherlands}


\begin{abstract}
Highly ambitious initiatives aspire to propel a miniature spacecraft to a neighboring star within a human generation, leveraging the radiation pressure of lasers for propulsion. One of the main challenges to achieving this enormous feat is to build a meter-scale, ultra-low mass lightsail with broadband reflectivity. In this work, we present the design and fabrication of such a lightsail composed of two distinct dielectric layers and patterned with a photonic crystal structure covering a 4" wafer. We overcome the crucial challenge of achieving broad band reflection of $>$70\% spanning over the full Doppler-shifted laser wavelength range during spacecraft acceleration, in combination with low total mass in the range of a few grams when scaled to meter size. Furthermore, we find new paths to reliably fabricate these sub-wavelength structures over macroscopic areas and then systematically characterize their optical performance, confirming their suitability for future lightsail applications. Our innovative device design and precise nanofabrication approaches represent a significant leap toward interstellar exploration.
\end{abstract}

\maketitle

\section*{Introduction}\label{sec1}

In the quest for interstellar exploration, the dream of propelling spacecraft to neighboring star systems has remained captivating for decades~\cite{marx1966interstellar,forward1984roundtrip,morris1988wormholes,lubin2016roadmap}. Within this ambitious pursuit, the Starshot Breakthrough Initiative has emerged as a leading forum to bring together scientists and engineers from many distinct fields and is driven by the goal of sending a spacecraft to Proxima Centauri, our nearest stellar neighbor, within the span of a human lifetime~\cite{starshot2019breakthrough}. At the heart of this mission lies the core concept -- a laser-driven lightsail. Once achieved, such lightsail technology would also enable many other ground-breaking space exploration missions, such as, exploring our own solar system within days rather than months or years, as well as harnessing the extraordinary imaging capabilities of the Solar Gravitational Lens (SGL) \cite{turyshev2020image,turyshev2022resolved}. Here, a lightsail is indispensable due to the SGL's optimal focal point being located over 548 astronomical units (AU) away from Earth. This vast distance necessitates a lightsail for delivering a camera to the SGL's focal point and precise positioning, enabling the capture of high-resolution images of exoplanets at 30 parsecs with a 10-kilometer-scale surface resolution -- unattainable with conventional spacecraft propulsion methods~\cite{turyshev2020direct}. The fundamental challenge confronting this endeavor requires the development of a lightweight spacecraft, which can be propelled by laser beams to extraordinary velocities, up to 20\% the speed of light~\cite{kulkarni2018relativistic,lubin2020path}. Unlike conventional propulsion systems~\cite{krejci2018space}, laser-driven lightsails rely on the radiation pressure force to achieve the immense speeds and acceleration required for interstellar space travel, which require the lightsail device to feature a large area, low mass, and broadband reflection. Several other crucial material requirements for designing a practical lightsail were thoroughly explored in previous work~\cite{atwater2018materials}, showing that the lightsail must have extreme optical, mechanical, and thermal properties to meet the constraints on mass and sail shape.

Previous works have simulated possible solutions with the practical constraints of making a lightsail. For example, it was demonstrated that by using inverse design and large-scale optimization, a lightweight broadband reflector for relativistic lightsail propulsion based on stacked photonic crystal slabs shows potential improvement in acceleration distance performance~\cite{jin2020inverse}. Similarly, it was also shown that optimized multilayer structures can enable ultralight spacecraft to sustain high acceleration while striking a balance between efficiency and weight~\cite{santi2022multilayers}. The choice of materials for these layers is crucial, ensuring high reflectance in the Doppler-shifted laser wavelength range. Although different works have carefully analyzed the design~\cite{ilic2018nanophotonic}, stability~\cite{myilswamy2020photonic,manchester2017stability}, and acceleration properties of a lightsail~\cite{macchi2009light}, no experimental realization of such a lightsail has been demonstrated to date, to the best of our knowledge.

\begin{figure*}
	\includegraphics[width = 1.0 \linewidth]{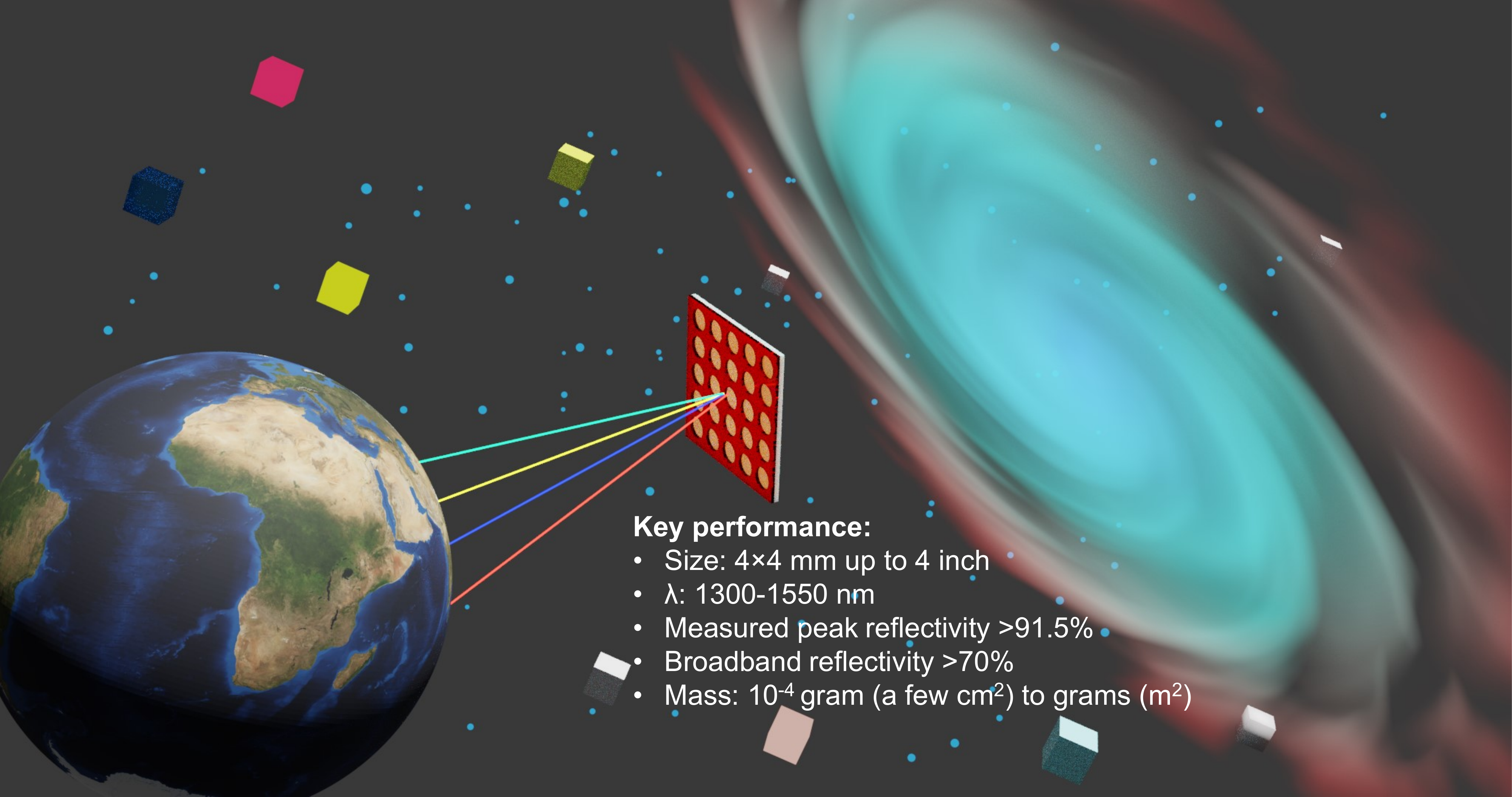}
	\caption{Illustration of the lightsail concept and the main performance parameters of the device realized in this work, including size, working wavelengths, reflection performance, and mass. The source image used to create the planet is taken from NASA's Visible Earth website~\cite{NASA_BlueMarble}.}
	\label{fig1}
\end{figure*}

One of the key challenges for a feasible lightsail design lies in constructing a broad-band reflector capable of covering the Doppler-shifted laser wavelength spectrum during the spacecraft's high-velocity acceleration phase. Earlier efforts to develop such a broadband reflector involved an InP 2D photonic crystal and Fano resonator~\cite{qiang2010design,boutami2006broadband}. However, these attempts were constrained to very small areas without complete substrate release. Moreover, in the later work, the added SiO$_2$ layer inevitably filled in the holes of the photonic crystal layer, resulting in a degraded reflection performance. For a more detailed discussion, we refer to~\cite{zhou2014progress}. In this work, we approach the material and design challenges associated with laser-driven lightsails, by pioneering a bilayer membrane structure using a silicon nitride photonic crystal atop a flat silicon layer, drawing inspiration from the high reflectivity exhibited by conventional photonic crystal devices~\cite{joannopoulos1997photonic,nair2010photonic} and the enhanced functionality and tunability offered by metasurface devices~\cite{hu2021review,ji2023recent}. Our novel approach encompasses the incorporation of a high refractive index layer beneath the photonic crystal membrane, resulting in a significant expansion of the reflectance spectrum. The bilayer configuration accomplishes broadband reflectivity (exceeding 70\%) spanning from 1300 to nearly 1550~nm, whereas a conventional single-layer photonic crystal only allows a reflectivity within the range of a few tens of nanometers. Furthermore, we establish a complete fabrication flow using silicon nitride on silicon-on-insulator (SOI) wafers, with new nanofabrication protocols that will allow scaling these structures from 4" wafers to the square-meter sizes required for future interstellar exploration.

\section*{Design and Simulation}\label{sec2}

In the following section, we would like to elaborate on the detailed design and simulation results of four different types of lightsail architectures and highlight the exceptional performance of our bi-layer membrane-based lightsails and their important role in advancing the frontiers of interstellar exploration. To design and simulate the reflection properties exhibited by distinct dielectric structures, we employ the electromagnetic field simulation software (CST Studio Suite). The refractive indices of our specific silicon (Si) and high-stress silicon nitride (SiN) films were first determined through ellipsometry measurements, with n$\und{Si}=3.4$ and n$\und{SiN} = 2.0$.

\begin{figure*}[ht]
    \centering
    \includegraphics[width=2\columnwidth]{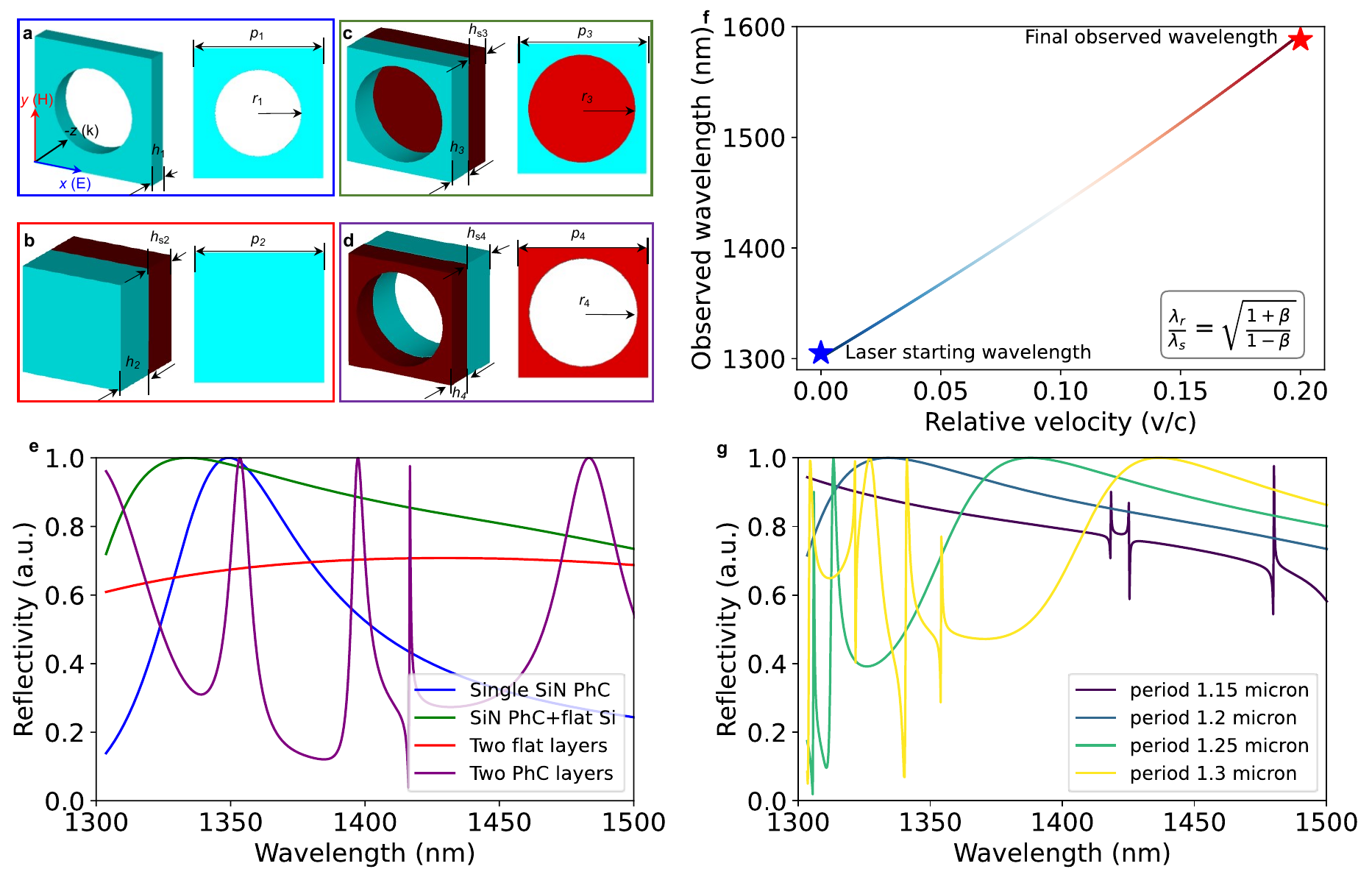}
    \caption{Simulated optical performance of different lightsail structures, and the Doppler effect for a fast-moving spacecraft. Figures (a)-(d) depict the side/top view of the single SiN photonic crystal, SiN/Si double flat layer membrane, SiN PhC/Si bilayer membrane, and SiN/Si double PhC structure, respectively. In (e) we compare each design's reflectivity and (f)  the Doppler effect of a moving lightsail from low speed up to 20\% of the speed of light, and (g) shows parameter sweep of the photonic crystal period values from 1.15 to 1.3~$\mu$m in the SiN PhC/Si bilayer membrane architecture to show how such design's reflectivity peak position can be flexibly engineered through modifications of the PhC layer parameter.}
    \label{fig2}
\end{figure*}

We first calculate the properties of a canonical single-layer photonic crystal periodic structure~\cite{guo2017integrated,gartner2018integrated}, visually represented in Fig.~\ref{fig2}a. Within this configuration, SiN serves as the dielectric material, with photonic crystal parameters $p_1=1200$~nm for the lattice constant, a thickness $h_1=400$~nm, and hole radius $r_1=500$~nm. Using the finite-difference time-domain (FDTD) algorithm, we conduct simulations with electromagnetic waves incident along the -z direction, and electric field polarization along the x direction, while imposing periodic boundary conditions. The simulation results, shown in Fig.~\ref{fig2}e as the blue curve, reveal that the reflectivity exceeds 99\% at its peak wavelength. As expected for this type of photonic crystal, we observe its reflectivity to exceed 70\% between 1330~nm and 1380~nm, corresponding to a bandwidth of 50~nm.

Due to their resonant character, photonic crystals are however inherently limited in their reflection bandwidth~\cite{gartner2018integrated}. To boost the reflection bandwidth and through inspiration by recent metasurface devices~\cite{ji2023recent}, we introduce a silicon optical impedance matching layer beneath the SiN. As a first step, we simulate how this affects the broadband reflection characteristics with both layers as simple continuous dielectric films (as shown in Fig.~\ref{fig2}b). The fabrication of such a structure is much simpler than a photonic crystal design and already allows us to see that the reflectivity exceeds 70\% within the wavelength range of 1385~nm to 1475~nm using the same lattice constant (red curve in Fig.~\ref{fig2}e). While the overall peak reflectivity is much lower compared to the one-layer PhC and the overall mass is greater, this approach points to a significant increase in the reflection bandwidth.

By combining the photonic crystal layer with a second high-refractive-index material underneath, we can now create a novel structure as illustrated in Fig.~\ref{fig2}c, which we call a \emph{meta-photonic crystal} (MPhC). The matching layer beneath the SiN, fabricated from silicon (depicted in red), is described by its thickness $hs_3=321$~nm, given by the specific SOI wafer we chose for this work. The new photonic crystal layer is now described by parameters p$_3$, h$_3$, and r$_3$. Using the same methodology, we simulate the MPhC's reflectivity, which surpasses 70\% for the entire 1300~nm to 1500~nm wavelength interval, with a peak reflectivity greater than 99\% (green curve in Fig.~\ref{fig2}e). This remarkable enhancement of the reflection bandwidth to around 200~nm, nearly quadrupling the bandwidth compared to the original structure while still maintaining an overall high reflectivity, holds significant promise for applications with broadband reflection requirements.

Finally, for completeness, we also study the design where a photonic crystal is fully etched through both layers, as shown in Fig.~\ref{fig2}d. The resulting reflectivity is plotted as the purple curves in Fig.~\ref{fig2}e. This design exhibits multiple resonance peaks in the 1300-1500~nm range, and we thus do not pursue it further as a design in our fabrication process.

\begin{figure*}[ht]
	\centering
	\includegraphics[width=2\columnwidth]{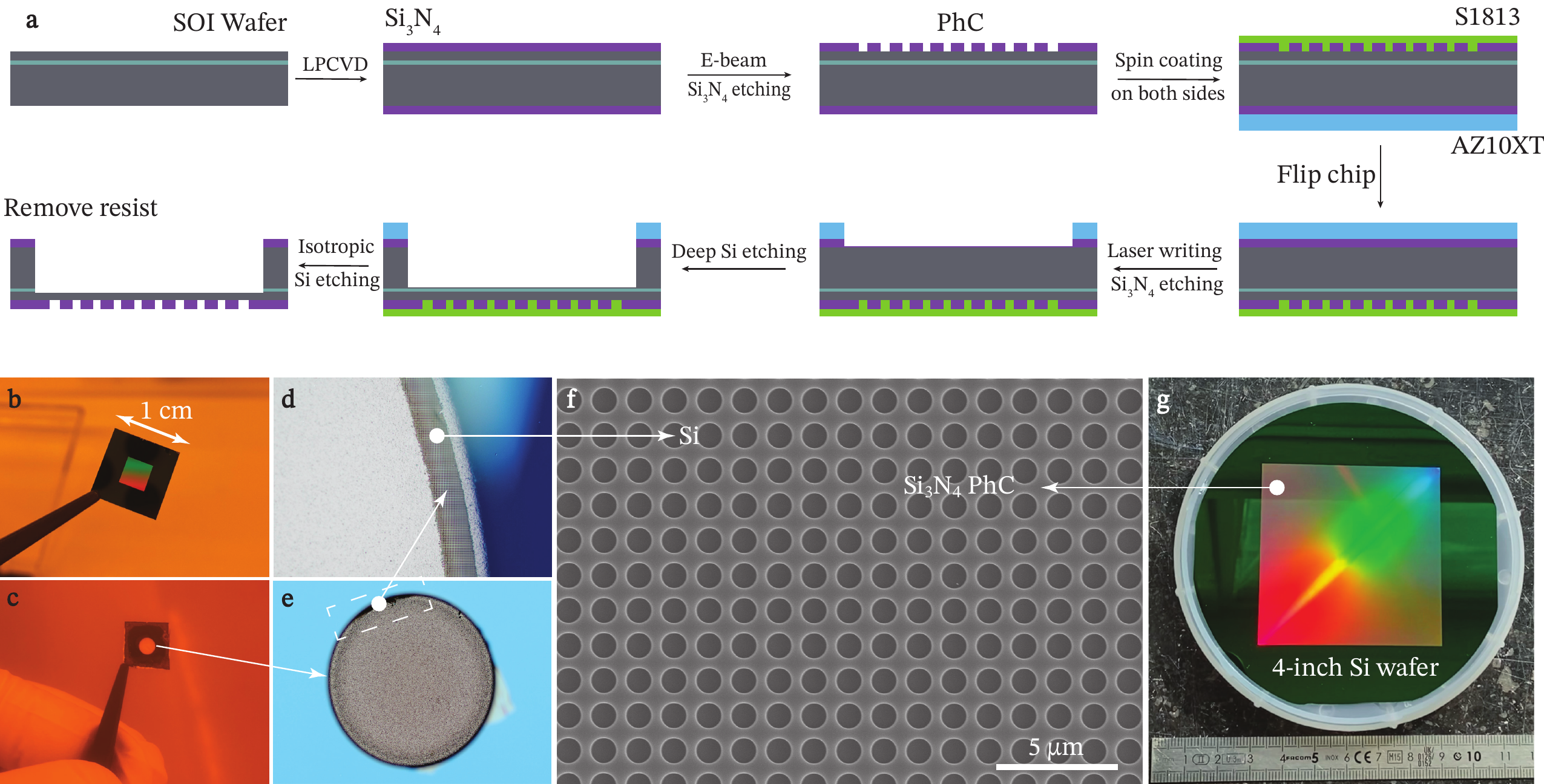}
	\caption{Fabrication flowchart of the lightsail starting from an SOI wafer and sample images at different fabrication steps. (a) illustrates the detailed meta-photonic crystal fabrication process, (b)-(e) show the optical images of the lightsail during the fabrication flow. (f) Scanning electron microscope (SEM) image of the lightsail device and (g) shows the full 4-inch size photonic crystal.}
	\label{fig3}
\end{figure*}

The new design of the meta-photonic crystal allows us to realize a broad-band reflector that seamlessly covers the entire wavelength range of a Doppler-shifted laser during the acceleration phase of the lightsail, while still maintaining its very low mass. To provide specific context, a laser with an initial wavelength of 1300~nm used to accelerate the lightsail to approximately 20\% the speed of light will experience a substantial shift, transitioning from 1300~nm to approximately 1550~nm, as depicted in Fig.~\ref{fig2}f. The Doppler shift is calculated using the relativistic Doppler effect formula~\cite{kaivola1985measurement,gordienko2004relativistic} (see inset in Fig.~\ref{fig2}f), where $\lambda_r$ is the lightsail's observed wavelength during acceleration, $\lambda_s$ is the laser wavelength on earth and $\beta=v/c$ is the speed of the lightsail normalized to the speed of light. It is worth mentioning that the MPhC lightsail's peak reflection position can be easily tuned by changing the photonic crystal layer's parameters, such as its periodicity. As shown in Fig.~\ref{fig2}g, the reflectivity peak changes from below 1300~nm to about 1420~nm by increasing the lattice constant from 1.15 to 1.3~$\mu$m. This offers high flexibility for designing a lightsail or other optical elements with adjustable optical responses, for example, meta-lenses~\cite{wang2018broadband}, on-chip integrated photonics devices~\cite{chang2023nanowire}, or free space optics~\cite{bayindir2002transmission}.

\section*{Device fabrication}\label{sec4}

Fabrication of small-scale photonic crystals is well established, and we use a similar approach as previous work~\cite{gartner2018integrated}. For a meta-photonic crystal device, as depicted in Fig.~\ref{fig3}a, the fabrication starts with a commercial SOI wafer with 321~nm thick silicon on a 394~nm thick buried oxide (BOX) layer, and a silicon handle wafer at the bottom, which is around 500~$\mu$m. A 400~nm high-stress stoichiometric silicon nitride (Si$_3$N$_4$) film is then deposited on both sides of the SOI wafer by low-pressure chemical vapor deposition (LPCVD) as described in~\cite{Guo2023}. The thickness of the silicon nitride layer and the silicon layer beneath are confirmed by ellipsometry measurements. After the LPCVD process, the 4-inch wafer is diced into 1$\times$1 centimeter chips. We pick individual chips and spin-coat their front side (MPhC side) with positive e-beam resists AR-P 6200 (CSAR 62) followed by electron beam lithography using the Raith EBPG 5200 with 100~kV acceleration voltage. After development (pentyl acetate for 1~min plus isopropyl alcohol for 1~min) and silicon nitride plasma etching using CH$_3$F/O$_2$ mixed gases~\cite{Guo2023}, the photonic crystal pattern, as simulated in the previous design and simulation section, is transferred to the silicon nitride layer. Then photoresist S1813 from Microresist is spin-coated on the chip front surface as a protection layer, and photoresist AZ10XT from MicroChemicals is spin-coated on the backside of the chip for next-step processing. The backside of the chip with AZ10XT resist is then exposed in a Heidelberg Instruments Laserwriter ($\mu$MLA) and developed to make a circular opening with a diameter of about 3~mm. Here, we use a circular-shaped opening, as it leads to a higher survival rate of the final bilayer membrane by reducing the stress concentration at the edge~\cite{gartner2018integrated}. The same plasma etch with CH$_3$F/O$_2$ mix gases is then used to first remove the 400~nm silicon nitride at the backside of the chip, and then a deep reactive ion silicon etching (Bosch process) is used to remove the 500~$\mu$m silicon handle layer. More details on this Bosch process are available in previous work~\cite{chang2021detecting}. When the silicon etching is finished, which can be confirmed by optical microscope observation from the color contrast between the silicon and silicon dioxide (cf.\ Fig.~\ref{fig3}d, where the SiO$_2$ is lighter in color than the Si), isotropic silicon etching with SF$_6$ gas is finally used to remove the BOX layer. Then the chip is carefully removed from the silicon carrier wafer and cleaned in hot acetone and isopropyl alcohol, leaving a clean silicon nitride/silicon bilayer membrane device as shown in Fig.~\ref{fig3}b. Significantly, the 400 nm high-stress silicon nitride and 321 nm silicon bilayer membrane exhibit notable mechanical resilience, resulting in a pronounced membrane survival rate during the whole fabrication process and upon final detachment from the silicon carrier wafer. Furthermore, the utilization of a thick and soft thermal compound adhesive between the chip and the silicon carrier wafer throughout the etching process, along with the prior application of spin-coated S1813 resist, ensures a high device yield.

To provide more details, we show a typical optical image of a $1\times1$~cm chip with $4\times4$~mm photonic crystal area (rainbow color) in Fig.~\ref{fig3}b. Similarly, Fig.~\ref{fig3}c shows the backside of the chip with a circular opening, which leads to higher device yield (exceeding 75\%) compared to rectangular openings, since sharp corners of the back opening result in membrane damage in our first few fabrication rounds. As mentioned before, when the deep silicon etching is nearly finished, the front side photonic crystal structure becomes visible through the edges of the backside opening, which can be seen under an optical microscope (Fig.~\ref{fig3}d and Fig.~\ref{fig3}e). We then stop the Bosch process and switch to isotropic silicon etch using SF$_6$ gas to remove the 394~nm BOX layer. The periodic photonic crystal structure is confirmed by scanning electron microscope (SEM) image, as presented in Fig.~\ref{fig3}f.

The dedicated fabrication process developed in this work can not only be used to produce centimeter-scale lightsails, but more importantly, it can directly be extended to scale up to larger sizes, such as a 4-inch wafer-scale fabrication process, as shown in Fig.~\ref{fig3}g. The rainbow-colored area in the middle of the wafer is the e-beam patterned and plasma-etched photonic crystal, where colors originate from scattered white light. To pattern such a large-area device with around $2.3\times10^9$ holes using e-beam lithography, we employ a custom-built program (called ``txl2gpf'') to generate GTX files in the Raith EBPG pattern data format. These files include sequences of beam positions, allowing the creation of single-shaped, high-resolution circles (with a subfield resolution of 0.08~nm). This approach contrasts with forming circles using numerous beam step-size resolution rectangles (ranging from 2 to 5~nm), leading to improved circle quality and significantly reduced e-beam writing time. The total e-beam writing time of such a 4-inch wafer sample is around 5–7 hours using a relatively large beam spot size of around 100~nm. We would like to note that a deep silicon etch is not performed on this sample, as dry etching introduces variation in the etch uniformity for such a large device. This challenge can, however, be easily overcome by employing wet-etching, e.g.\ through a potassium hydroxide (KOH) wet etching process~\cite{sato1998characterization} for both silicon and SiO$_2$ to get a large-scale lightsail device. In order to scale the fabrication further, larger scale wafers can be used or multiple 4-inch sized lightsail devices could be connected together to assemble a square-meter-sized lightsail, on which different sensors, receivers, and transmitters can be attached and then delivered to deep space through laser propulsion.

\section*{Optical Measurements}\label{sec5}

\begin{figure}[ht]
    \centering
    \includegraphics[width=0.9\columnwidth]{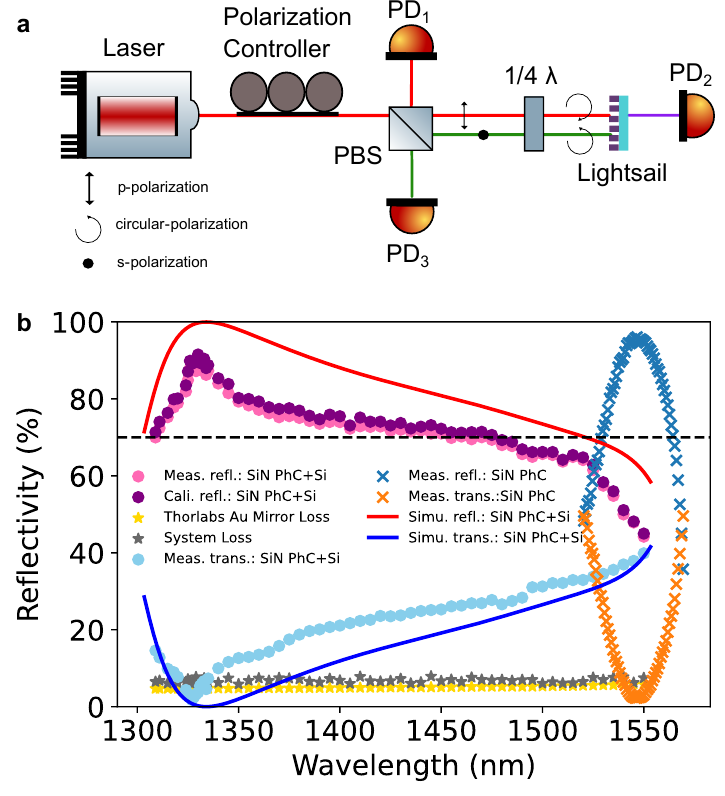}
    \caption{Characterization of the fabricated lightsail samples, including (a) optical measurement setup, and (b) measurements of both PhC and MPhC samples. See text for more details}
    \label{fig4}
\end{figure}

To test the fabricated samples and confirm our simulations through measurements, we use the setup depicted in Fig.~\ref{fig4}a. A tunable laser emitting light within the wavelength range of 1280-1600~nm serves as the source. Following the laser, a polarization controller (PC) is positioned, which we use to minimize the signal detected at photodetector 1 (PD$_1$) after the polarizing beamsplitter (PBS), ensuring the input light is linearly polarized (p-polarization). Subsequently, the transmitted light passes through a quarter wavelength wave plate, converting its polarization from linear to circular before reaching the lightsail sample. The light transmitted through the device is captured by PD$_2$ and recorded as P$\und{trans}$. Conversely, the reflected light is measured on PD$_3$ to obtain the reflected power P$\und{refl}$.

The optical power detected at PD$_1$ is negligible ($<$0.5\%) compared to the initial input power from the laser P$\und{in}$. Thus, we compute the transmission and reflection of the lightsail samples using the ratios P$\und{trans}$/P$\und{in}$ and P$\und{refl}$/P$\und{in}$, respectively. Due to small inherent losses in the optical setup, arising from factors such as misalignment of laser beams or imperfections in optical components, a small percentage of the total input laser power is lost. Consequently, P$\und{refl}$+P$\und{trans}$ is slightly lower than P$\und{in}$. To quantify these losses, we calibrate the system using a commercial gold mirror~\cite{ThorlabsWebsite}. Figure~\ref{fig4}b illustrates the system loss (depicted by gray stars) between 1300 and 1550~nm when employing a gold mirror to measure the discrepancy between P$\und{refl}$+P$\und{trans}$ and P$\und{in}$. The overall system losses range between 5\% to 7\%, which is in great part attributable to the gold mirror itself, for which we use the datasheet from the manufacturer (gold stars in Fig.~\ref{fig4}b).

Upon calibration, we evaluate the first batch of fabricated MPhC lightsail chips, each approximately 4$\times$4~mm in size. As indicated by the pink dots in Figure~\ref{fig4}b, the MPhC sample (denoted as ‘SiN PhC + Si’) achieves broadband reflection from 1300 to nearly 1550~nm (70\% reflectivity threshold is indicated by the black dashed line). The peak reflectivity exceeds 90\% at 1330~nm, with a bandwidth of nearly 200~nm. The corresponding transmission is denoted by the light-blue dots. Accounting for the aforementioned system losses, the calibrated reflectivity (depicted by purple dots) exhibits a similarly broad performance, reaching a peak reflectivity of over 91.5\% at 1330~nm. The red and blue curves represent the simulated reflection and transmission of the lightsail, respectively. Discrepancies between simulations and measurements are attributed to fabrication deviations or imperfections in the measurement setup.

In comparison, the typical single-layer silicon nitride photonic crystal, referred to as ‘SiN PhC,’ achieves $>$70\% reflectivity only between 1525 and 1575~nm, with a narrow bandwidth of 50~nm. This bandwidth is significantly narrower than that of the bilayer lightsail design, highlighting the superior performance of the MPhC devices.

\section*{Theoretical Analysis}\label{sec6}

To understand why the SiN PhC/Si bilayer membrane structure exhibits a broader reflection bandwidth compared to a single-layer SiN photonic crystal, we perform further simulations based on the theory of multi-layered media equivalent wave impedance~\cite{stanciulescu1980optical,pedrotti2017introduction}. The results are presented in Fig.~\ref{fig5} below.

\begin{figure}[ht]
    \centering
    \includegraphics[width=\columnwidth]{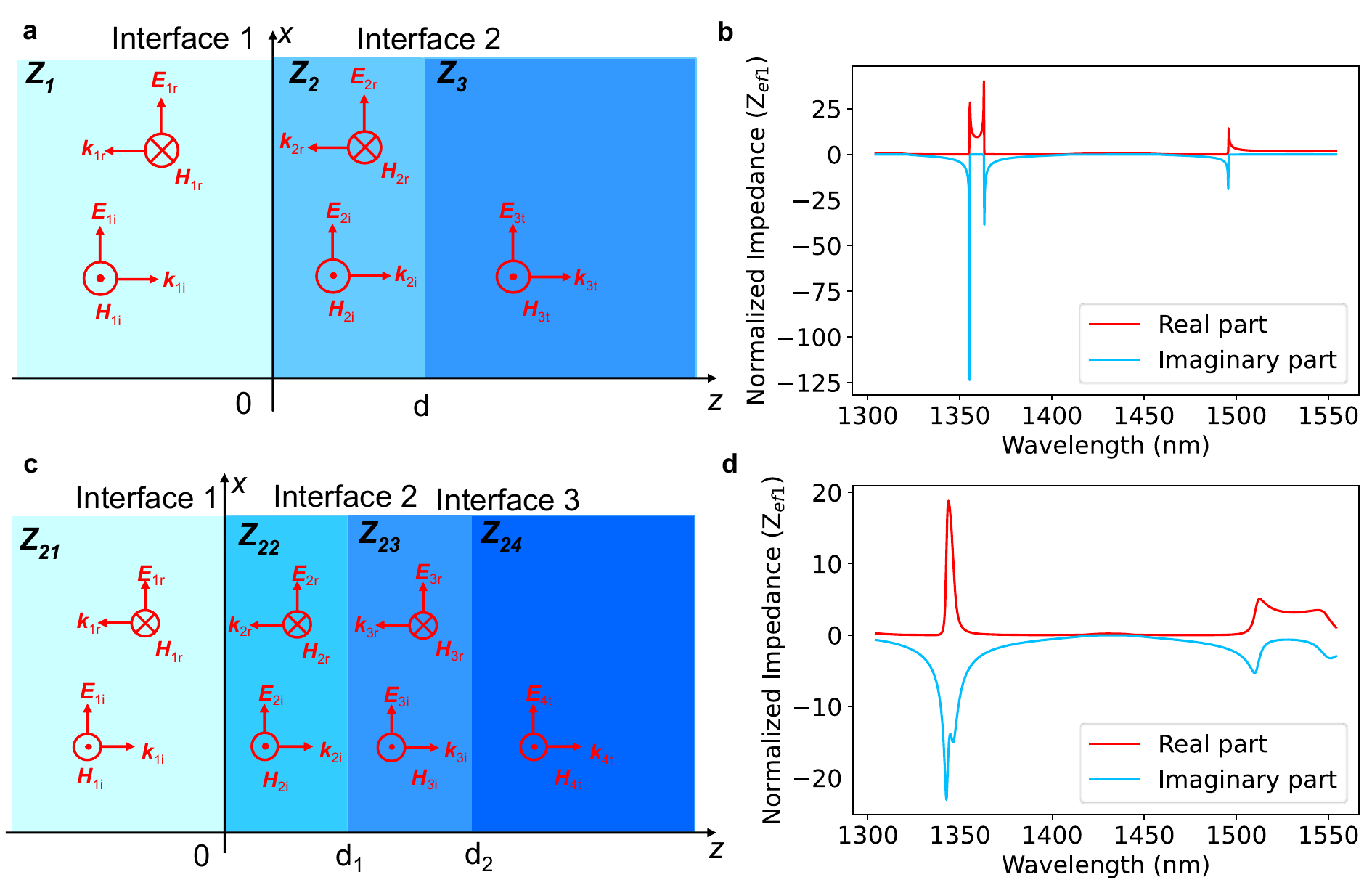}
    \caption{Multi-layered media equivalent wave impedance simulation and analysis of both SiN photonic crystal and SiN PhC/Si bilayer membrane architecture.}
    \label{fig5}
\end{figure}

In Fig.~\ref{fig5}a, going from left to right, we have three media labeled 1, 2, and 3, with their normalized impedance denoted as Z$_1$, Z$_2$, and Z$_3$. These three distinct lossless media have parallel interfaces at $z=0$ and $z=d$, with the thickness of media 2 being denoted as $d$. When an electromagnetic wave vertically impinges from media 1 along the positive $z$-axis, reflections, and transmissions occur at the interfaces at $z=0$ and $z=d$. Consequently, there are incident waves propagating in the $+z$ direction and reflected waves propagating in the $-z$ direction in both media 1 and media 2, while only transmitted waves propagate in media 3. The reflection coefficient at the interface $z=0$ is given by

\begin{equation}
	R_1 = \frac{Z\und{ef23} - Z_1}{Z\und{ef23} + Z_1}, \label{eq1}
\end{equation}

\noindent where Z$\und{ef23}$ represents the equivalent wave impedance of media 2 and media 3 at $z=0$, and

\begin{equation}
	Z_{\und{ef23}} = Z_2 \frac{Z_3 + jZ_2 \tan(\beta_{2}d)}{Z_2 + jZ_3 \tan(\beta_{2}d)}, \label{eq2}
\end{equation}

\noindent where $j$ is the imaginary part, and $\beta_2 = \omega \sqrt{\epsilon \mu}$ is the propagation constant of the electromagnetic wave in media 2. In the case of the single-layer SiN photonic crystal structure, the SiN photonic crystal serves as media 2, with air layers both in the forward (media 1) and backward (media 3) directions, resulting in Z$_1$ = Z$_3$ = Z$_0$. Based on the reflectance coefficient of the SiN photonic crystal in Fig.~\ref{fig2}g and Eq.~(\ref{eq1}), we can compute the normalized equivalent impedance Z$\und{ef23}$, as depicted in Fig.~\ref{fig5}b.

Subsequently, we consider the SiN PhC/Si bilayer membrane model. As this structure consists of two media layers, we equivalently model it as a four-layered media structure, as shown in Fig.~\ref{fig5}c. Silicon represents media 2, and the SiN photonic crystal corresponds to media 3, with their normalized impedance denoted as $Z_{22} = Z_0 \sqrt{\frac{\mu_r}{\epsilon_r}}$ and Z$_{23}$, respectively. The forward (media 1) and backward (media 4) directions in this case both consist of air layers, resulting in Z$_{21}$ = Z$_{24}$ = Z$_{0}$. At the interface $z=d_1$ between media 3 and media 4, the equivalent wave impedance Z$\und{ef2}$ = Z$\und{ef23}$, as previously determined in Fig.~\ref{fig2}b. Subsequently, according to Eq.~(\ref{eq2}), we calculate the equivalent wave impedance at $z=0$ for media 2, 3, and 4 as

\begin{equation}
	Z\und{ef1} = Z_{22} \frac{Z\und{ef2} + jZ_{22} \tan(\beta_{2}{d_1})}{Z_{22} + jZ\und{ef2} \tan(\beta_{2}{d_1})}. \label{eq3}
\end{equation}

\noindent Finally, by applying Eq.~\eqref{eq1}, we calculate the reflection coefficients as

\begin{equation}
	R_2 = \frac{Z\und{ef1} - Z_0}{Z\und{ef1} + Z_0} \label{eq4}.
\end{equation}

\noindent The theoretical calculation results of the reflection coefficient for the designed reflector are presented in Fig.~\ref{fig5}d. It can be seen that the real and imaginary parts of the MPhC structure's optical impedance are significantly modified around 1500~nm, thus its high reflectance is extended into longer wavelength compared to a single-layer SiN device. Therefore, by introducing an additional Si layer and appropriately designing its thickness and photonic crystal parameters, we can achieve high-efficiency and wideband reflection.

\section*{Conclusion and Future Perspectives}\label{sec13}

In summary, our study pioneers the development of a photonic crystal/metasurface bilayer structure for laser-driven lightsails, overcoming critical challenges in the quest for interstellar travel. Our innovative design, featuring a silicon nitride photonic crystal with a thin silicon membrane, achieves outstanding experimentally demonstrated high ($>$91.5\% at 1330~nm) and broadband reflectivity (exceeding 70\% from 1300 to 1500~nm). This broad reflection spectrum is essential for accommodating Doppler-shifted laser wavelengths during lightsail acceleration. The dedicated fabrication process involves precise techniques, including wafer-scale nanophotonic structure patterning and deep silicon etching. Rigorous optical measurements and theoretical analyses confirm our lightsail's performance, marking a significant advancement in lightsail design and fabrication and verifying its capabilities for future interstellar exploration.

Looking ahead, the scope of lightsail research extends to larger dimensions on the meter scale, achievable through methods like wet chemical etching, to obtain large, freestanding reflective surfaces. Moreover, our study, while focusing on the SiN/Si bilayer structure, does not limit itself to this material system. Bilayers with significant refractive index contrast, such as SiN/SiC or normal SiN/silicon-rich silicon nitride, can also be employed to design broadband reflectors, potentially minimizing any residual absorption. Additionally, this high-reflectivity, wideband reflection structure has the potential to also play a vital role in various other fields, including optoelectronic devices~\cite{wang2021chip,esmaeil2021superconducting}, integrated optics~\cite{yue2017highly,bao2020demand}, metamaterials and devices~\cite{zhao2019integrating,padilla2022imaging}, opening up new possibilities for future applications.

\section*{Acknowledgments}
We gratefully acknowledge Johannes W.N.\ Los and Matthijs H.J.\ de Jong for their support in building the optical measurement setup, as well as Carel Heerkens for their help in preparing the SOI wafer. This work is financially supported by the European Research Council (ERC CoG Q-ECHOS, 101001005), and by the Netherlands Organization for Scientific Research (NWO/OCW), as part of the Frontiers of Nanoscience program, as well as through a Vrij Programma (680-92-18-04) grant and the Breakthrough Starshot Foundation.

\section*{Data availability}
Source data for the figures will be made available on Zenodo.

\section*{Competing interests}
The authors declare no competing interests.

\bibliography{sn-bibliography}

\begin{thebibliography}{45}%
\makeatletter
\providecommand \@ifxundefined [1]{%
 \@ifx{#1\undefined}
}%
\providecommand \@ifnum [1]{%
 \ifnum #1\expandafter \@firstoftwo
 \else \expandafter \@secondoftwo
 \fi
}%
\providecommand \@ifx [1]{%
 \ifx #1\expandafter \@firstoftwo
 \else \expandafter \@secondoftwo
 \fi
}%
\providecommand \natexlab [1]{#1}%
\providecommand \enquote  [1]{``#1''}%
\providecommand \bibnamefont  [1]{#1}%
\providecommand \bibfnamefont [1]{#1}%
\providecommand \citenamefont [1]{#1}%
\providecommand \href@noop [0]{\@secondoftwo}%
\providecommand \href [0]{\begingroup \@sanitize@url \@href}%
\providecommand \@href[1]{\@@startlink{#1}\@@href}%
\providecommand \@@href[1]{\endgroup#1\@@endlink}%
\providecommand \@sanitize@url [0]{\catcode `\\12\catcode `\$12\catcode
  `\&12\catcode `\#12\catcode `\^12\catcode `\_12\catcode `\%12\relax}%
\providecommand \@@startlink[1]{}%
\providecommand \@@endlink[0]{}%
\providecommand \url  [0]{\begingroup\@sanitize@url \@url }%
\providecommand \@url [1]{\endgroup\@href {#1}{\urlprefix }}%
\providecommand \urlprefix  [0]{URL }%
\providecommand \Eprint [0]{\href }%
\providecommand \doibase [0]{https://doi.org/}%
\providecommand \selectlanguage [0]{\@gobble}%
\providecommand \bibinfo  [0]{\@secondoftwo}%
\providecommand \bibfield  [0]{\@secondoftwo}%
\providecommand \translation [1]{[#1]}%
\providecommand \BibitemOpen [0]{}%
\providecommand \bibitemStop [0]{}%
\providecommand \bibitemNoStop [0]{.\EOS\space}%
\providecommand \EOS [0]{\spacefactor3000\relax}%
\providecommand \BibitemShut  [1]{\csname bibitem#1\endcsname}%
\let\auto@bib@innerbib\@empty
\bibitem [{\citenamefont {Marx}(1966)}]{marx1966interstellar}%
  \BibitemOpen
  \bibfield  {author} {\bibinfo {author} {\bibfnamefont {G.}~\bibnamefont
  {Marx}},\ }\bibfield  {title} {\bibinfo {title} {Interstellar vehicle
  propelled by terrestrial laser beam},\ }\href
  {https://doi.org/10.1038/211022a0} {\bibfield  {journal} {\bibinfo  {journal}
  {Nature}\ }\textbf {\bibinfo {volume} {211}},\ \bibinfo {pages} {22}
  (\bibinfo {year} {1966})}\BibitemShut {NoStop}%
\bibitem [{\citenamefont {Forward}(1984)}]{forward1984roundtrip}%
  \BibitemOpen
  \bibfield  {author} {\bibinfo {author} {\bibfnamefont {R.~L.}\ \bibnamefont
  {Forward}},\ }\bibfield  {title} {\bibinfo {title} {Roundtrip interstellar
  travel using laser-pushed lightsails},\ }\href
  {https://doi.org/10.2514/3.8632} {\bibfield  {journal} {\bibinfo  {journal}
  {J. Spacecr. Rockets}\ }\textbf {\bibinfo {volume} {21}},\ \bibinfo {pages}
  {187} (\bibinfo {year} {1984})}\BibitemShut {NoStop}%
\bibitem [{\citenamefont {Morris}\ and\ \citenamefont
  {Thorne}(1988)}]{morris1988wormholes}%
  \BibitemOpen
  \bibfield  {author} {\bibinfo {author} {\bibfnamefont {M.~S.}\ \bibnamefont
  {Morris}}\ and\ \bibinfo {author} {\bibfnamefont {K.~S.}\ \bibnamefont
  {Thorne}},\ }\bibfield  {title} {\bibinfo {title} {Wormholes in spacetime and
  their use for interstellar travel: A tool for teaching general relativity},\
  }\href {https://doi.org/10.1119/1.15620} {\bibfield  {journal} {\bibinfo
  {journal} {Am. J. Phys.}\ }\textbf {\bibinfo {volume} {56}},\ \bibinfo
  {pages} {395} (\bibinfo {year} {1988})}\BibitemShut {NoStop}%
\bibitem [{\citenamefont {Lubin}(2016)}]{lubin2016roadmap}%
  \BibitemOpen
  \bibfield  {author} {\bibinfo {author} {\bibfnamefont {P.}~\bibnamefont
  {Lubin}},\ }\bibfield  {title} {\bibinfo {title} {A roadmap to interstellar
  flight},\ }\href {https://doi.org/10.48550/arXiv.1604.01356} {\bibfield
  {journal} {\bibinfo  {journal} {J. Br. Interplanet. Soc.}\ }\textbf {\bibinfo
  {volume} {69}},\ \bibinfo {pages} {40} (\bibinfo {year} {2016})}\BibitemShut
  {NoStop}%
\bibitem [{sta()}]{starshot2019breakthrough}%
  \BibitemOpen
  \href {https://breakthroughinitiatives.org/initiative/3} {\bibinfo {title}
  {{Starshot Breakthrough Initiative}}}\BibitemShut {NoStop}%
\bibitem [{\citenamefont {Turyshev}\ and\ \citenamefont
  {Toth}(2020)}]{turyshev2020image}%
  \BibitemOpen
  \bibfield  {author} {\bibinfo {author} {\bibfnamefont {S.~G.}\ \bibnamefont
  {Turyshev}}\ and\ \bibinfo {author} {\bibfnamefont {V.~T.}\ \bibnamefont
  {Toth}},\ }\bibfield  {title} {\bibinfo {title} {Image formation process with
  the solar gravitational lens},\ }\href
  {https://doi.org/10.1103/physrevd.101.044048} {\bibfield  {journal} {\bibinfo
   {journal} {Phys. Rev. D}\ }\textbf {\bibinfo {volume} {101}},\ \bibinfo
  {pages} {044048} (\bibinfo {year} {2020})}\BibitemShut {NoStop}%
\bibitem [{\citenamefont {Turyshev}\ and\ \citenamefont
  {Toth}(2022)}]{turyshev2022resolved}%
  \BibitemOpen
  \bibfield  {author} {\bibinfo {author} {\bibfnamefont {S.~G.}\ \bibnamefont
  {Turyshev}}\ and\ \bibinfo {author} {\bibfnamefont {V.~T.}\ \bibnamefont
  {Toth}},\ }\bibfield  {title} {\bibinfo {title} {Resolved imaging of
  exoplanets with the solar gravitational lens},\ }\href
  {https://doi.org/10.1093/mnras/stac2130} {\bibfield  {journal} {\bibinfo
  {journal} {Mon. Notices Royal Astron. Soc.}\ }\textbf {\bibinfo {volume}
  {515}},\ \bibinfo {pages} {6122} (\bibinfo {year} {2022})}\BibitemShut
  {NoStop}%
\bibitem [{\citenamefont {Turyshev}\ \emph {et~al.}(2020)\citenamefont
  {Turyshev}, \citenamefont {Shao}, \citenamefont {Toth}, \citenamefont
  {Friedman}, \citenamefont {Alkalai}, \citenamefont {Mawet}, \citenamefont
  {Shen}, \citenamefont {Swain}, \citenamefont {Zhou}, \citenamefont
  {Helvajian} \emph {et~al.}}]{turyshev2020direct}%
  \BibitemOpen
  \bibfield  {author} {\bibinfo {author} {\bibfnamefont {S.~G.}\ \bibnamefont
  {Turyshev}}, \bibinfo {author} {\bibfnamefont {M.}~\bibnamefont {Shao}},
  \bibinfo {author} {\bibfnamefont {V.~T.}\ \bibnamefont {Toth}}, \bibinfo
  {author} {\bibfnamefont {L.~D.}\ \bibnamefont {Friedman}}, \bibinfo {author}
  {\bibfnamefont {L.}~\bibnamefont {Alkalai}}, \bibinfo {author} {\bibfnamefont
  {D.}~\bibnamefont {Mawet}}, \bibinfo {author} {\bibfnamefont
  {J.}~\bibnamefont {Shen}}, \bibinfo {author} {\bibfnamefont {M.~R.}\
  \bibnamefont {Swain}}, \bibinfo {author} {\bibfnamefont {H.}~\bibnamefont
  {Zhou}}, \bibinfo {author} {\bibfnamefont {H.}~\bibnamefont {Helvajian}},
  \emph {et~al.},\ }\bibfield  {title} {\bibinfo {title} {Direct multipixel
  imaging and spectroscopy of an exoplanet with a solar gravity lens mission},\
  }\bibfield  {journal} {\bibinfo  {journal} {arXiv:2002.11871}\ }\href
  {https://doi.org/10.48550/arXiv.2002.11871} {10.48550/arXiv.2002.11871}
  (\bibinfo {year} {2020})\BibitemShut {NoStop}%
\bibitem [{\citenamefont {Kulkarni}\ \emph {et~al.}(2018)\citenamefont
  {Kulkarni}, \citenamefont {Lubin},\ and\ \citenamefont
  {Zhang}}]{kulkarni2018relativistic}%
  \BibitemOpen
  \bibfield  {author} {\bibinfo {author} {\bibfnamefont {N.}~\bibnamefont
  {Kulkarni}}, \bibinfo {author} {\bibfnamefont {P.}~\bibnamefont {Lubin}},\
  and\ \bibinfo {author} {\bibfnamefont {Q.}~\bibnamefont {Zhang}},\ }\bibfield
   {title} {\bibinfo {title} {Relativistic spacecraft propelled by directed
  energy},\ }\href {https://doi.org/10.3847/1538-3881/aaafd2} {\bibfield
  {journal} {\bibinfo  {journal} {Astron. J.}\ }\textbf {\bibinfo {volume}
  {155}},\ \bibinfo {pages} {155} (\bibinfo {year} {2018})}\BibitemShut
  {NoStop}%
\bibitem [{\citenamefont {Lubin}\ and\ \citenamefont
  {Hettel}(2020)}]{lubin2020path}%
  \BibitemOpen
  \bibfield  {author} {\bibinfo {author} {\bibfnamefont {P.}~\bibnamefont
  {Lubin}}\ and\ \bibinfo {author} {\bibfnamefont {W.}~\bibnamefont {Hettel}},\
  }\bibfield  {title} {\bibinfo {title} {The path to interstellar flight},\
  }\href {https://doi.org/10.5281/zenodo.3747263} {\bibfield  {journal}
  {\bibinfo  {journal} {Acta Futura}\ }\textbf {\bibinfo {volume} {12}},\
  \bibinfo {pages} {9} (\bibinfo {year} {2020})}\BibitemShut {NoStop}%
\bibitem [{\citenamefont {Krejci}\ and\ \citenamefont
  {Lozano}(2018)}]{krejci2018space}%
  \BibitemOpen
  \bibfield  {author} {\bibinfo {author} {\bibfnamefont {D.}~\bibnamefont
  {Krejci}}\ and\ \bibinfo {author} {\bibfnamefont {P.}~\bibnamefont
  {Lozano}},\ }\bibfield  {title} {\bibinfo {title} {Space propulsion
  technology for small spacecraft},\ }\href
  {https://doi.org/10.1109/jproc.2017.2778747} {\bibfield  {journal} {\bibinfo
  {journal} {Proc. IEEE}\ }\textbf {\bibinfo {volume} {106}},\ \bibinfo {pages}
  {362} (\bibinfo {year} {2018})}\BibitemShut {NoStop}%
\bibitem [{\citenamefont {Atwater}\ \emph {et~al.}(2018)\citenamefont
  {Atwater}, \citenamefont {Davoyan}, \citenamefont {Ilic}, \citenamefont
  {Jariwala}, \citenamefont {Sherrott}, \citenamefont {Went}, \citenamefont
  {Whitney},\ and\ \citenamefont {Wong}}]{atwater2018materials}%
  \BibitemOpen
  \bibfield  {author} {\bibinfo {author} {\bibfnamefont {H.~A.}\ \bibnamefont
  {Atwater}}, \bibinfo {author} {\bibfnamefont {A.~R.}\ \bibnamefont
  {Davoyan}}, \bibinfo {author} {\bibfnamefont {O.}~\bibnamefont {Ilic}},
  \bibinfo {author} {\bibfnamefont {D.}~\bibnamefont {Jariwala}}, \bibinfo
  {author} {\bibfnamefont {M.~C.}\ \bibnamefont {Sherrott}}, \bibinfo {author}
  {\bibfnamefont {C.~M.}\ \bibnamefont {Went}}, \bibinfo {author}
  {\bibfnamefont {W.~S.}\ \bibnamefont {Whitney}},\ and\ \bibinfo {author}
  {\bibfnamefont {J.}~\bibnamefont {Wong}},\ }\bibfield  {title} {\bibinfo
  {title} {Materials challenges for the starshot lightsail},\ }\href
  {https://doi.org/10.1038/s41563-018-0075-8} {\bibfield  {journal} {\bibinfo
  {journal} {Nat. Mater.}\ }\textbf {\bibinfo {volume} {17}},\ \bibinfo {pages}
  {861} (\bibinfo {year} {2018})}\BibitemShut {NoStop}%
\bibitem [{\citenamefont {Jin}\ \emph {et~al.}(2020)\citenamefont {Jin},
  \citenamefont {Li}, \citenamefont {Orenstein},\ and\ \citenamefont
  {Fan}}]{jin2020inverse}%
  \BibitemOpen
  \bibfield  {author} {\bibinfo {author} {\bibfnamefont {W.}~\bibnamefont
  {Jin}}, \bibinfo {author} {\bibfnamefont {W.}~\bibnamefont {Li}}, \bibinfo
  {author} {\bibfnamefont {M.}~\bibnamefont {Orenstein}},\ and\ \bibinfo
  {author} {\bibfnamefont {S.}~\bibnamefont {Fan}},\ }\bibfield  {title}
  {\bibinfo {title} {Inverse design of lightweight broadband reflector for
  relativistic lightsail propulsion},\ }\href
  {https://doi.org/10.1021/acsphotonics.0c00768.s001} {\bibfield  {journal}
  {\bibinfo  {journal} {ACS Photonics}\ }\textbf {\bibinfo {volume} {7}},\
  \bibinfo {pages} {2350} (\bibinfo {year} {2020})}\BibitemShut {NoStop}%
\bibitem [{\citenamefont {Santi}\ \emph {et~al.}(2022)\citenamefont {Santi},
  \citenamefont {Favaro}, \citenamefont {Corso}, \citenamefont {Lubin},
  \citenamefont {Bazzan}, \citenamefont {Ragazzoni}, \citenamefont {Garoli},\
  and\ \citenamefont {Pelizzo}}]{santi2022multilayers}%
  \BibitemOpen
  \bibfield  {author} {\bibinfo {author} {\bibfnamefont {G.}~\bibnamefont
  {Santi}}, \bibinfo {author} {\bibfnamefont {G.}~\bibnamefont {Favaro}},
  \bibinfo {author} {\bibfnamefont {A.~J.}\ \bibnamefont {Corso}}, \bibinfo
  {author} {\bibfnamefont {P.}~\bibnamefont {Lubin}}, \bibinfo {author}
  {\bibfnamefont {M.}~\bibnamefont {Bazzan}}, \bibinfo {author} {\bibfnamefont
  {R.}~\bibnamefont {Ragazzoni}}, \bibinfo {author} {\bibfnamefont
  {D.}~\bibnamefont {Garoli}},\ and\ \bibinfo {author} {\bibfnamefont {M.~G.}\
  \bibnamefont {Pelizzo}},\ }\bibfield  {title} {\bibinfo {title} {Multilayers
  for directed energy accelerated lightsails},\ }\href
  {https://doi.org/10.1038/s43246-022-00240-8} {\bibfield  {journal} {\bibinfo
  {journal} {Commun. Mater.}\ }\textbf {\bibinfo {volume} {3}},\ \bibinfo
  {pages} {16} (\bibinfo {year} {2022})}\BibitemShut {NoStop}%
\bibitem [{\citenamefont {Ilic}\ \emph {et~al.}(2018)\citenamefont {Ilic},
  \citenamefont {Went},\ and\ \citenamefont {Atwater}}]{ilic2018nanophotonic}%
  \BibitemOpen
  \bibfield  {author} {\bibinfo {author} {\bibfnamefont {O.}~\bibnamefont
  {Ilic}}, \bibinfo {author} {\bibfnamefont {C.~M.}\ \bibnamefont {Went}},\
  and\ \bibinfo {author} {\bibfnamefont {H.~A.}\ \bibnamefont {Atwater}},\
  }\bibfield  {title} {\bibinfo {title} {Nanophotonic heterostructures for
  efficient propulsion and radiative cooling of relativistic light sails},\
  }\href {https://doi.org/10.1021/acs.nanolett.8b02035.s001} {\bibfield
  {journal} {\bibinfo  {journal} {Nano Lett.}\ }\textbf {\bibinfo {volume}
  {18}},\ \bibinfo {pages} {5583} (\bibinfo {year} {2018})}\BibitemShut
  {NoStop}%
\bibitem [{\citenamefont {Myilswamy}\ \emph {et~al.}(2020)\citenamefont
  {Myilswamy}, \citenamefont {Krishnan},\ and\ \citenamefont
  {Povinelli}}]{myilswamy2020photonic}%
  \BibitemOpen
  \bibfield  {author} {\bibinfo {author} {\bibfnamefont {K.~V.}\ \bibnamefont
  {Myilswamy}}, \bibinfo {author} {\bibfnamefont {A.}~\bibnamefont
  {Krishnan}},\ and\ \bibinfo {author} {\bibfnamefont {M.~L.}\ \bibnamefont
  {Povinelli}},\ }\bibfield  {title} {\bibinfo {title} {Photonic crystal
  lightsail with nonlinear reflectivity for increased stability},\ }\href
  {https://doi.org/10.1364/oe.387687} {\bibfield  {journal} {\bibinfo
  {journal} {Opt. Express}\ }\textbf {\bibinfo {volume} {28}},\ \bibinfo
  {pages} {8223} (\bibinfo {year} {2020})}\BibitemShut {NoStop}%
\bibitem [{\citenamefont {Manchester}\ and\ \citenamefont
  {Loeb}(2017)}]{manchester2017stability}%
  \BibitemOpen
  \bibfield  {author} {\bibinfo {author} {\bibfnamefont {Z.}~\bibnamefont
  {Manchester}}\ and\ \bibinfo {author} {\bibfnamefont {A.}~\bibnamefont
  {Loeb}},\ }\bibfield  {title} {\bibinfo {title} {Stability of a light sail
  riding on a laser beam},\ }\href {https://doi.org/10.3847/2041-8213/aa619b}
  {\bibfield  {journal} {\bibinfo  {journal} {Astrophys. J. Lett.}\ }\textbf
  {\bibinfo {volume} {837}},\ \bibinfo {pages} {L20} (\bibinfo {year}
  {2017})}\BibitemShut {NoStop}%
\bibitem [{\citenamefont {Macchi}\ \emph {et~al.}(2009)\citenamefont {Macchi},
  \citenamefont {Veghini},\ and\ \citenamefont {Pegoraro}}]{macchi2009light}%
  \BibitemOpen
  \bibfield  {author} {\bibinfo {author} {\bibfnamefont {A.}~\bibnamefont
  {Macchi}}, \bibinfo {author} {\bibfnamefont {S.}~\bibnamefont {Veghini}},\
  and\ \bibinfo {author} {\bibfnamefont {F.}~\bibnamefont {Pegoraro}},\
  }\bibfield  {title} {\bibinfo {title} {“light sail” acceleration
  reexamined},\ }\href {https://doi.org/10.1103/physrevlett.103.085003}
  {\bibfield  {journal} {\bibinfo  {journal} {Phys. Rev. Lett.}\ }\textbf
  {\bibinfo {volume} {103}},\ \bibinfo {pages} {085003} (\bibinfo {year}
  {2009})}\BibitemShut {NoStop}%
\bibitem [{NAS()}]{NASA_BlueMarble}%
  \BibitemOpen
  \href {https://visibleearth.nasa.gov/collection/1484/blue-marble} {\bibinfo
  {title} {Blue marble, {NASA Visible Earth}}}\BibitemShut {NoStop}%
\bibitem [{\citenamefont {Qiang}\ \emph {et~al.}(2010)\citenamefont {Qiang},
  \citenamefont {Yang}, \citenamefont {Chuwongin}, \citenamefont {Zhao},
  \citenamefont {Ma},\ and\ \citenamefont {Zhou}}]{qiang2010design}%
  \BibitemOpen
  \bibfield  {author} {\bibinfo {author} {\bibfnamefont {Z.}~\bibnamefont
  {Qiang}}, \bibinfo {author} {\bibfnamefont {H.}~\bibnamefont {Yang}},
  \bibinfo {author} {\bibfnamefont {S.}~\bibnamefont {Chuwongin}}, \bibinfo
  {author} {\bibfnamefont {D.}~\bibnamefont {Zhao}}, \bibinfo {author}
  {\bibfnamefont {Z.}~\bibnamefont {Ma}},\ and\ \bibinfo {author}
  {\bibfnamefont {W.}~\bibnamefont {Zhou}},\ }\bibfield  {title} {\bibinfo
  {title} {Design of fano broadband reflectors on soi},\ }\href
  {https://doi.org/https://doi.org/10.1109/lpt.2010.2050471} {\bibfield
  {journal} {\bibinfo  {journal} {IEEE Photonics Technology Letters}\ }\textbf
  {\bibinfo {volume} {22}},\ \bibinfo {pages} {1108} (\bibinfo {year}
  {2010})}\BibitemShut {NoStop}%
\bibitem [{\citenamefont {Boutami}\ \emph {et~al.}(2006)\citenamefont
  {Boutami}, \citenamefont {Bakir}, \citenamefont {Hattori}, \citenamefont
  {Letartre}, \citenamefont {Leclercq}, \citenamefont {Rojo-Romeo},
  \citenamefont {Garrigues}, \citenamefont {Seassal},\ and\ \citenamefont
  {Viktorovitch}}]{boutami2006broadband}%
  \BibitemOpen
  \bibfield  {author} {\bibinfo {author} {\bibfnamefont {S.}~\bibnamefont
  {Boutami}}, \bibinfo {author} {\bibfnamefont {B.~B.}\ \bibnamefont {Bakir}},
  \bibinfo {author} {\bibfnamefont {H.}~\bibnamefont {Hattori}}, \bibinfo
  {author} {\bibfnamefont {X.}~\bibnamefont {Letartre}}, \bibinfo {author}
  {\bibfnamefont {J.-L.}\ \bibnamefont {Leclercq}}, \bibinfo {author}
  {\bibfnamefont {P.}~\bibnamefont {Rojo-Romeo}}, \bibinfo {author}
  {\bibfnamefont {M.}~\bibnamefont {Garrigues}}, \bibinfo {author}
  {\bibfnamefont {C.}~\bibnamefont {Seassal}},\ and\ \bibinfo {author}
  {\bibfnamefont {P.}~\bibnamefont {Viktorovitch}},\ }\bibfield  {title}
  {\bibinfo {title} {Broadband and compact 2-d photonic crystal reflectors with
  controllable polarization dependence},\ }\href
  {https://doi.org/https://doi.org/10.1109/lpt.2006.871833} {\bibfield
  {journal} {\bibinfo  {journal} {IEEE Photonics Technology Letters}\ }\textbf
  {\bibinfo {volume} {18}},\ \bibinfo {pages} {835} (\bibinfo {year}
  {2006})}\BibitemShut {NoStop}%
\bibitem [{\citenamefont {Zhou}\ \emph {et~al.}(2014)\citenamefont {Zhou},
  \citenamefont {Zhao}, \citenamefont {Shuai}, \citenamefont {Yang},
  \citenamefont {Chuwongin}, \citenamefont {Chadha}, \citenamefont {Seo},
  \citenamefont {Wang}, \citenamefont {Liu}, \citenamefont {Ma} \emph
  {et~al.}}]{zhou2014progress}%
  \BibitemOpen
  \bibfield  {author} {\bibinfo {author} {\bibfnamefont {W.}~\bibnamefont
  {Zhou}}, \bibinfo {author} {\bibfnamefont {D.}~\bibnamefont {Zhao}}, \bibinfo
  {author} {\bibfnamefont {Y.-C.}\ \bibnamefont {Shuai}}, \bibinfo {author}
  {\bibfnamefont {H.}~\bibnamefont {Yang}}, \bibinfo {author} {\bibfnamefont
  {S.}~\bibnamefont {Chuwongin}}, \bibinfo {author} {\bibfnamefont
  {A.}~\bibnamefont {Chadha}}, \bibinfo {author} {\bibfnamefont {J.-H.}\
  \bibnamefont {Seo}}, \bibinfo {author} {\bibfnamefont {K.~X.}\ \bibnamefont
  {Wang}}, \bibinfo {author} {\bibfnamefont {V.}~\bibnamefont {Liu}}, \bibinfo
  {author} {\bibfnamefont {Z.}~\bibnamefont {Ma}}, \emph {et~al.},\ }\bibfield
  {title} {\bibinfo {title} {Progress in 2d photonic crystal fano resonance
  photonics},\ }\href
  {https://doi.org/https://doi.org/10.1016/j.pquantelec.2014.01.001} {\bibfield
   {journal} {\bibinfo  {journal} {Progress in Quantum Electronics}\ }\textbf
  {\bibinfo {volume} {38}},\ \bibinfo {pages} {1} (\bibinfo {year}
  {2014})}\BibitemShut {NoStop}%
\bibitem [{\citenamefont {Joannopoulos}\ \emph {et~al.}(1997)\citenamefont
  {Joannopoulos}, \citenamefont {Villeneuve},\ and\ \citenamefont
  {Fan}}]{joannopoulos1997photonic}%
  \BibitemOpen
  \bibfield  {author} {\bibinfo {author} {\bibfnamefont {J.~D.}\ \bibnamefont
  {Joannopoulos}}, \bibinfo {author} {\bibfnamefont {P.~R.}\ \bibnamefont
  {Villeneuve}},\ and\ \bibinfo {author} {\bibfnamefont {S.}~\bibnamefont
  {Fan}},\ }\bibfield  {title} {\bibinfo {title} {Photonic crystals},\ }\href
  {https://doi.org/10.1016/S0038-1098(96)00716-8} {\bibfield  {journal}
  {\bibinfo  {journal} {Solid State Commun.}\ }\textbf {\bibinfo {volume}
  {102}},\ \bibinfo {pages} {165} (\bibinfo {year} {1997})}\BibitemShut
  {NoStop}%
\bibitem [{\citenamefont {Nair}\ and\ \citenamefont
  {Vijaya}(2010)}]{nair2010photonic}%
  \BibitemOpen
  \bibfield  {author} {\bibinfo {author} {\bibfnamefont {R.~V.}\ \bibnamefont
  {Nair}}\ and\ \bibinfo {author} {\bibfnamefont {R.}~\bibnamefont {Vijaya}},\
  }\bibfield  {title} {\bibinfo {title} {Photonic crystal sensors: An
  overview},\ }\href {https://doi.org/10.1016/j.pquantelec.2010.01.001}
  {\bibfield  {journal} {\bibinfo  {journal} {Prog. Quantum. Electron.}\
  }\textbf {\bibinfo {volume} {34}},\ \bibinfo {pages} {89} (\bibinfo {year}
  {2010})}\BibitemShut {NoStop}%
\bibitem [{\citenamefont {Hu}\ \emph {et~al.}(2021)\citenamefont {Hu},
  \citenamefont {Bandyopadhyay}, \citenamefont {Liu},\ and\ \citenamefont
  {Shao}}]{hu2021review}%
  \BibitemOpen
  \bibfield  {author} {\bibinfo {author} {\bibfnamefont {J.}~\bibnamefont
  {Hu}}, \bibinfo {author} {\bibfnamefont {S.}~\bibnamefont {Bandyopadhyay}},
  \bibinfo {author} {\bibfnamefont {Y.-h.}\ \bibnamefont {Liu}},\ and\ \bibinfo
  {author} {\bibfnamefont {L.-y.}\ \bibnamefont {Shao}},\ }\bibfield  {title}
  {\bibinfo {title} {A review on metasurface: from principle to smart
  metadevices},\ }\href {https://doi.org/10.3389/fphy.2020.586087} {\bibfield
  {journal} {\bibinfo  {journal} {Front. Phys.}\ }\textbf {\bibinfo {volume}
  {8}},\ \bibinfo {pages} {586087} (\bibinfo {year} {2021})}\BibitemShut
  {NoStop}%
\bibitem [{\citenamefont {Ji}\ \emph {et~al.}(2023)\citenamefont {Ji},
  \citenamefont {Chang}, \citenamefont {Xu}, \citenamefont {Gao}, \citenamefont
  {Gr{\"o}blacher}, \citenamefont {Urbach},\ and\ \citenamefont
  {Adam}}]{ji2023recent}%
  \BibitemOpen
  \bibfield  {author} {\bibinfo {author} {\bibfnamefont {W.}~\bibnamefont
  {Ji}}, \bibinfo {author} {\bibfnamefont {J.}~\bibnamefont {Chang}}, \bibinfo
  {author} {\bibfnamefont {H.-X.}\ \bibnamefont {Xu}}, \bibinfo {author}
  {\bibfnamefont {J.~R.}\ \bibnamefont {Gao}}, \bibinfo {author} {\bibfnamefont
  {S.}~\bibnamefont {Gr{\"o}blacher}}, \bibinfo {author} {\bibfnamefont
  {H.~P.}\ \bibnamefont {Urbach}},\ and\ \bibinfo {author} {\bibfnamefont
  {A.~J.}\ \bibnamefont {Adam}},\ }\bibfield  {title} {\bibinfo {title} {Recent
  advances in metasurface design and quantum optics applications with machine
  learning, physics-informed neural networks, and topology optimization
  methods},\ }\href {https://doi.org/10.1038/s41377-023-01218-y} {\bibfield
  {journal} {\bibinfo  {journal} {Light Sci. Appl.}\ }\textbf {\bibinfo
  {volume} {12}},\ \bibinfo {pages} {169} (\bibinfo {year} {2023})}\BibitemShut
  {NoStop}%
\bibitem [{\citenamefont {Guo}\ \emph {et~al.}(2017)\citenamefont {Guo},
  \citenamefont {Norte},\ and\ \citenamefont
  {Gr{\"o}blacher}}]{guo2017integrated}%
  \BibitemOpen
  \bibfield  {author} {\bibinfo {author} {\bibfnamefont {J.}~\bibnamefont
  {Guo}}, \bibinfo {author} {\bibfnamefont {R.~A.}\ \bibnamefont {Norte}},\
  and\ \bibinfo {author} {\bibfnamefont {S.}~\bibnamefont {Gr{\"o}blacher}},\
  }\bibfield  {title} {\bibinfo {title} {Integrated optical force sensors using
  focusing photonic crystal arrays},\ }\href
  {https://doi.org/10.1364/oe.25.009196} {\bibfield  {journal} {\bibinfo
  {journal} {Opt. Express}\ }\textbf {\bibinfo {volume} {25}},\ \bibinfo
  {pages} {9196} (\bibinfo {year} {2017})}\BibitemShut {NoStop}%
\bibitem [{\citenamefont {G\"{a}rtner}\ \emph {et~al.}(2018)\citenamefont
  {G\"{a}rtner}, \citenamefont {Moura}, \citenamefont {Haaxman}, \citenamefont
  {Norte},\ and\ \citenamefont {Gr\"{o}blacher}}]{gartner2018integrated}%
  \BibitemOpen
  \bibfield  {author} {\bibinfo {author} {\bibfnamefont {C.}~\bibnamefont
  {G\"{a}rtner}}, \bibinfo {author} {\bibfnamefont {J.~P.}\ \bibnamefont
  {Moura}}, \bibinfo {author} {\bibfnamefont {W.}~\bibnamefont {Haaxman}},
  \bibinfo {author} {\bibfnamefont {R.~A.}\ \bibnamefont {Norte}},\ and\
  \bibinfo {author} {\bibfnamefont {S.}~\bibnamefont {Gr\"{o}blacher}},\
  }\bibfield  {title} {\bibinfo {title} {Integrated optomechanical arrays of
  two high reflectivity sin membranes},\ }\href
  {https://doi.org/10.1021/acs.nanolett.8b03240} {\bibfield  {journal}
  {\bibinfo  {journal} {Nano Lett.}\ }\textbf {\bibinfo {volume} {18}},\
  \bibinfo {pages} {7171} (\bibinfo {year} {2018})}\BibitemShut {NoStop}%
\bibitem [{\citenamefont {Kaivola}\ \emph {et~al.}(1985)\citenamefont
  {Kaivola}, \citenamefont {Poulsen}, \citenamefont {Riis},\ and\ \citenamefont
  {Lee}}]{kaivola1985measurement}%
  \BibitemOpen
  \bibfield  {author} {\bibinfo {author} {\bibfnamefont {M.}~\bibnamefont
  {Kaivola}}, \bibinfo {author} {\bibfnamefont {O.}~\bibnamefont {Poulsen}},
  \bibinfo {author} {\bibfnamefont {E.}~\bibnamefont {Riis}},\ and\ \bibinfo
  {author} {\bibfnamefont {S.~A.}\ \bibnamefont {Lee}},\ }\bibfield  {title}
  {\bibinfo {title} {Measurement of the relativistic doppler shift in neon},\
  }\href {https://doi.org/10.1103/physrevlett.54.255} {\bibfield  {journal}
  {\bibinfo  {journal} {Phys. Rev. Lett.}\ }\textbf {\bibinfo {volume} {54}},\
  \bibinfo {pages} {255} (\bibinfo {year} {1985})}\BibitemShut {NoStop}%
\bibitem [{\citenamefont {Gordienko}\ \emph {et~al.}(2004)\citenamefont
  {Gordienko}, \citenamefont {Pukhov}, \citenamefont {Shorokhov},\ and\
  \citenamefont {Baeva}}]{gordienko2004relativistic}%
  \BibitemOpen
  \bibfield  {author} {\bibinfo {author} {\bibfnamefont {S.}~\bibnamefont
  {Gordienko}}, \bibinfo {author} {\bibfnamefont {A.}~\bibnamefont {Pukhov}},
  \bibinfo {author} {\bibfnamefont {O.}~\bibnamefont {Shorokhov}},\ and\
  \bibinfo {author} {\bibfnamefont {T.}~\bibnamefont {Baeva}},\ }\bibfield
  {title} {\bibinfo {title} {Relativistic doppler effect: Universal spectra and
  zeptosecond pulses},\ }\href {https://doi.org/10.1103/physrevlett.93.115002}
  {\bibfield  {journal} {\bibinfo  {journal} {Phys. Rev. Lett.}\ }\textbf
  {\bibinfo {volume} {93}},\ \bibinfo {pages} {115002} (\bibinfo {year}
  {2004})}\BibitemShut {NoStop}%
\bibitem [{\citenamefont {Wang}\ \emph {et~al.}(2018)\citenamefont {Wang},
  \citenamefont {Wu}, \citenamefont {Su}, \citenamefont {Lai}, \citenamefont
  {Chen}, \citenamefont {Kuo}, \citenamefont {Chen}, \citenamefont {Chen},
  \citenamefont {Huang}, \citenamefont {Wang} \emph
  {et~al.}}]{wang2018broadband}%
  \BibitemOpen
  \bibfield  {author} {\bibinfo {author} {\bibfnamefont {S.}~\bibnamefont
  {Wang}}, \bibinfo {author} {\bibfnamefont {P.~C.}\ \bibnamefont {Wu}},
  \bibinfo {author} {\bibfnamefont {V.-C.}\ \bibnamefont {Su}}, \bibinfo
  {author} {\bibfnamefont {Y.-C.}\ \bibnamefont {Lai}}, \bibinfo {author}
  {\bibfnamefont {M.-K.}\ \bibnamefont {Chen}}, \bibinfo {author}
  {\bibfnamefont {H.~Y.}\ \bibnamefont {Kuo}}, \bibinfo {author} {\bibfnamefont
  {B.~H.}\ \bibnamefont {Chen}}, \bibinfo {author} {\bibfnamefont {Y.~H.}\
  \bibnamefont {Chen}}, \bibinfo {author} {\bibfnamefont {T.-T.}\ \bibnamefont
  {Huang}}, \bibinfo {author} {\bibfnamefont {J.-H.}\ \bibnamefont {Wang}},
  \emph {et~al.},\ }\bibfield  {title} {\bibinfo {title} {A broadband
  achromatic metalens in the visible},\ }\href
  {https://doi.org/10.1038/s41565-017-0052-4} {\bibfield  {journal} {\bibinfo
  {journal} {Nat. Nanotechnol.}\ }\textbf {\bibinfo {volume} {13}},\ \bibinfo
  {pages} {227} (\bibinfo {year} {2018})}\BibitemShut {NoStop}%
\bibitem [{\citenamefont {Chang}\ \emph {et~al.}(2023)\citenamefont {Chang},
  \citenamefont {Gao}, \citenamefont {Esmaeil~Zadeh}, \citenamefont
  {Elshaari},\ and\ \citenamefont {Zwiller}}]{chang2023nanowire}%
  \BibitemOpen
  \bibfield  {author} {\bibinfo {author} {\bibfnamefont {J.}~\bibnamefont
  {Chang}}, \bibinfo {author} {\bibfnamefont {J.}~\bibnamefont {Gao}}, \bibinfo
  {author} {\bibfnamefont {I.}~\bibnamefont {Esmaeil~Zadeh}}, \bibinfo {author}
  {\bibfnamefont {A.~W.}\ \bibnamefont {Elshaari}},\ and\ \bibinfo {author}
  {\bibfnamefont {V.}~\bibnamefont {Zwiller}},\ }\bibfield  {title} {\bibinfo
  {title} {Nanowire-based integrated photonics for quantum information and
  quantum sensing},\ }\href {https://doi.org/10.1515/nanoph-2022-0652}
  {\bibfield  {journal} {\bibinfo  {journal} {Nanophotonics}\ }\textbf
  {\bibinfo {volume} {12}},\ \bibinfo {pages} {339} (\bibinfo {year}
  {2023})}\BibitemShut {NoStop}%
\bibitem [{\citenamefont {Bayindir}\ \emph {et~al.}(2002)\citenamefont
  {Bayindir}, \citenamefont {Aydin}, \citenamefont {Ozbay}, \citenamefont
  {Marko{\v{s}}},\ and\ \citenamefont {Soukoulis}}]{bayindir2002transmission}%
  \BibitemOpen
  \bibfield  {author} {\bibinfo {author} {\bibfnamefont {M.}~\bibnamefont
  {Bayindir}}, \bibinfo {author} {\bibfnamefont {K.}~\bibnamefont {Aydin}},
  \bibinfo {author} {\bibfnamefont {E.}~\bibnamefont {Ozbay}}, \bibinfo
  {author} {\bibfnamefont {P.}~\bibnamefont {Marko{\v{s}}}},\ and\ \bibinfo
  {author} {\bibfnamefont {C.}~\bibnamefont {Soukoulis}},\ }\bibfield  {title}
  {\bibinfo {title} {Transmission properties of composite metamaterials in free
  space},\ }\href {https://doi.org/10.1063/1.1492009} {\bibfield  {journal}
  {\bibinfo  {journal} {Appl. Phys. Lett.}\ }\textbf {\bibinfo {volume} {81}},\
  \bibinfo {pages} {120} (\bibinfo {year} {2002})}\BibitemShut {NoStop}%
\bibitem [{\citenamefont {Guo}\ \emph {et~al.}(2023)\citenamefont {Guo},
  \citenamefont {Chang}, \citenamefont {Yao},\ and\ \citenamefont
  {Gr{\"o}blacher}}]{Guo2023}%
  \BibitemOpen
  \bibfield  {author} {\bibinfo {author} {\bibfnamefont {J.}~\bibnamefont
  {Guo}}, \bibinfo {author} {\bibfnamefont {J.}~\bibnamefont {Chang}}, \bibinfo
  {author} {\bibfnamefont {X.}~\bibnamefont {Yao}},\ and\ \bibinfo {author}
  {\bibfnamefont {S.}~\bibnamefont {Gr{\"o}blacher}},\ }\bibfield  {title}
  {\bibinfo {title} {Active-feedback quantum control of an integrated
  low-frequency mechanical resonator},\ }\href
  {https://doi.org/10.1038/s41467-023-40442-3} {\bibfield  {journal} {\bibinfo
  {journal} {Nat. Commun.}\ }\textbf {\bibinfo {volume} {14}},\ \bibinfo
  {pages} {4721} (\bibinfo {year} {2023})}\BibitemShut {NoStop}%
\bibitem [{\citenamefont {Chang}\ \emph {et~al.}(2021)\citenamefont {Chang},
  \citenamefont {Los}, \citenamefont {Tenorio-Pearl}, \citenamefont {Noordzij},
  \citenamefont {Gourgues}, \citenamefont {Guardiani}, \citenamefont {Zichi},
  \citenamefont {Pereira}, \citenamefont {Urbach}, \citenamefont {Zwiller}
  \emph {et~al.}}]{chang2021detecting}%
  \BibitemOpen
  \bibfield  {author} {\bibinfo {author} {\bibfnamefont {J.}~\bibnamefont
  {Chang}}, \bibinfo {author} {\bibfnamefont {J.}~\bibnamefont {Los}}, \bibinfo
  {author} {\bibfnamefont {J.}~\bibnamefont {Tenorio-Pearl}}, \bibinfo {author}
  {\bibfnamefont {N.}~\bibnamefont {Noordzij}}, \bibinfo {author}
  {\bibfnamefont {R.}~\bibnamefont {Gourgues}}, \bibinfo {author}
  {\bibfnamefont {A.}~\bibnamefont {Guardiani}}, \bibinfo {author}
  {\bibfnamefont {J.}~\bibnamefont {Zichi}}, \bibinfo {author} {\bibfnamefont
  {S.}~\bibnamefont {Pereira}}, \bibinfo {author} {\bibfnamefont
  {H.}~\bibnamefont {Urbach}}, \bibinfo {author} {\bibfnamefont
  {V.}~\bibnamefont {Zwiller}}, \emph {et~al.},\ }\bibfield  {title} {\bibinfo
  {title} {Detecting telecom single photons with 99.5- 2.07+ 0.5\% system
  detection efficiency and high time resolution},\ }\href
  {https://doi.org/10.1063/5.0039772} {\bibfield  {journal} {\bibinfo
  {journal} {APL Photonics}\ }\textbf {\bibinfo {volume} {6}},\ \bibinfo
  {pages} {036114} (\bibinfo {year} {2021})}\BibitemShut {NoStop}%
\bibitem [{\citenamefont {Sato}\ \emph {et~al.}(1998)\citenamefont {Sato},
  \citenamefont {Shikida}, \citenamefont {Matsushima}, \citenamefont
  {Yamashiro}, \citenamefont {Asaumi}, \citenamefont {Iriye},\ and\
  \citenamefont {Yamamoto}}]{sato1998characterization}%
  \BibitemOpen
  \bibfield  {author} {\bibinfo {author} {\bibfnamefont {K.}~\bibnamefont
  {Sato}}, \bibinfo {author} {\bibfnamefont {M.}~\bibnamefont {Shikida}},
  \bibinfo {author} {\bibfnamefont {Y.}~\bibnamefont {Matsushima}}, \bibinfo
  {author} {\bibfnamefont {T.}~\bibnamefont {Yamashiro}}, \bibinfo {author}
  {\bibfnamefont {K.}~\bibnamefont {Asaumi}}, \bibinfo {author} {\bibfnamefont
  {Y.}~\bibnamefont {Iriye}},\ and\ \bibinfo {author} {\bibfnamefont
  {M.}~\bibnamefont {Yamamoto}},\ }\bibfield  {title} {\bibinfo {title}
  {Characterization of orientation-dependent etching properties of
  single-crystal silicon: effects of koh concentration},\ }\href
  {https://doi.org/10.1016/s0924-4247(97)01658-0} {\bibfield  {journal}
  {\bibinfo  {journal} {Sens. Actuator A Phys.}\ }\textbf {\bibinfo {volume}
  {64}},\ \bibinfo {pages} {87} (\bibinfo {year} {1998})}\BibitemShut {NoStop}%
\bibitem [{\citenamefont {{Thorlabs}}()}]{ThorlabsWebsite}%
  \BibitemOpen
  \bibfield  {author} {\bibinfo {author} {\bibnamefont {{Thorlabs}}},\ }\href
  {https://www.thorlabs.com/newgrouppage9.cfm?objectgroup_id=744} {\bibinfo
  {title} {Protected gold mirrors}}\BibitemShut {NoStop}%
\bibitem [{\citenamefont {Stanciulescu}\ \emph {et~al.}(1980)\citenamefont
  {Stanciulescu}, \citenamefont {Bobulescu}, \citenamefont {Surmeian},
  \citenamefont {Popescu}, \citenamefont {Popescu},\ and\ \citenamefont
  {Collins}}]{stanciulescu1980optical}%
  \BibitemOpen
  \bibfield  {author} {\bibinfo {author} {\bibfnamefont {C.}~\bibnamefont
  {Stanciulescu}}, \bibinfo {author} {\bibfnamefont {R.}~\bibnamefont
  {Bobulescu}}, \bibinfo {author} {\bibfnamefont {A.}~\bibnamefont {Surmeian}},
  \bibinfo {author} {\bibfnamefont {D.}~\bibnamefont {Popescu}}, \bibinfo
  {author} {\bibfnamefont {I.}~\bibnamefont {Popescu}},\ and\ \bibinfo {author}
  {\bibfnamefont {C.}~\bibnamefont {Collins}},\ }\bibfield  {title} {\bibinfo
  {title} {Optical impedance spectroscopy},\ }\href
  {https://doi.org/https://doi.org/10.1063/1.91785} {\bibfield  {journal}
  {\bibinfo  {journal} {Applied Physics Letters}\ }\textbf {\bibinfo {volume}
  {37}},\ \bibinfo {pages} {888} (\bibinfo {year} {1980})}\BibitemShut
  {NoStop}%
\bibitem [{\citenamefont {Pedrotti}\ \emph {et~al.}(2017)\citenamefont
  {Pedrotti}, \citenamefont {Pedrotti},\ and\ \citenamefont
  {Pedrotti}}]{pedrotti2017introduction}%
  \BibitemOpen
  \bibfield  {author} {\bibinfo {author} {\bibfnamefont {F.~L.}\ \bibnamefont
  {Pedrotti}}, \bibinfo {author} {\bibfnamefont {L.~M.}\ \bibnamefont
  {Pedrotti}},\ and\ \bibinfo {author} {\bibfnamefont {L.~S.}\ \bibnamefont
  {Pedrotti}},\ }\href {https://doi.org/10.1017/9781108552493} {\emph {\bibinfo
  {title} {Introduction to Optics}}}\ (\bibinfo  {publisher} {Cambridge
  University Press},\ \bibinfo {year} {2017})\BibitemShut {NoStop}%
\bibitem [{\citenamefont {Wang}\ \emph {et~al.}(2021)\citenamefont {Wang},
  \citenamefont {Ni}, \citenamefont {Xie}, \citenamefont {Kan}, \citenamefont
  {Chen}, \citenamefont {Fu}, \citenamefont {Deng}, \citenamefont {Jin},
  \citenamefont {Chen}, \citenamefont {Lee} \emph {et~al.}}]{wang2021chip}%
  \BibitemOpen
  \bibfield  {author} {\bibinfo {author} {\bibfnamefont {Q.-H.}\ \bibnamefont
  {Wang}}, \bibinfo {author} {\bibfnamefont {P.-N.}\ \bibnamefont {Ni}},
  \bibinfo {author} {\bibfnamefont {Y.-Y.}\ \bibnamefont {Xie}}, \bibinfo
  {author} {\bibfnamefont {Q.}~\bibnamefont {Kan}}, \bibinfo {author}
  {\bibfnamefont {P.-P.}\ \bibnamefont {Chen}}, \bibinfo {author}
  {\bibfnamefont {P.}~\bibnamefont {Fu}}, \bibinfo {author} {\bibfnamefont
  {J.}~\bibnamefont {Deng}}, \bibinfo {author} {\bibfnamefont {T.-L.}\
  \bibnamefont {Jin}}, \bibinfo {author} {\bibfnamefont {H.-D.}\ \bibnamefont
  {Chen}}, \bibinfo {author} {\bibfnamefont {H.~W.~H.}\ \bibnamefont {Lee}},
  \emph {et~al.},\ }\bibfield  {title} {\bibinfo {title} {On-chip generation of
  structured light based on metasurface optoelectronic integration},\ }\href
  {https://doi.org/10.1002/lpor.202000385} {\bibfield  {journal} {\bibinfo
  {journal} {Laser Photonics Rev.}\ }\textbf {\bibinfo {volume} {15}},\
  \bibinfo {pages} {2000385} (\bibinfo {year} {2021})}\BibitemShut {NoStop}%
\bibitem [{\citenamefont {Esmaeil~Zadeh}\ \emph {et~al.}(2021)\citenamefont
  {Esmaeil~Zadeh}, \citenamefont {Chang}, \citenamefont {Los}, \citenamefont
  {Gyger}, \citenamefont {Elshaari}, \citenamefont {Steinhauer}, \citenamefont
  {Dorenbos},\ and\ \citenamefont {Zwiller}}]{esmaeil2021superconducting}%
  \BibitemOpen
  \bibfield  {author} {\bibinfo {author} {\bibfnamefont {I.}~\bibnamefont
  {Esmaeil~Zadeh}}, \bibinfo {author} {\bibfnamefont {J.}~\bibnamefont
  {Chang}}, \bibinfo {author} {\bibfnamefont {J.~W.}\ \bibnamefont {Los}},
  \bibinfo {author} {\bibfnamefont {S.}~\bibnamefont {Gyger}}, \bibinfo
  {author} {\bibfnamefont {A.~W.}\ \bibnamefont {Elshaari}}, \bibinfo {author}
  {\bibfnamefont {S.}~\bibnamefont {Steinhauer}}, \bibinfo {author}
  {\bibfnamefont {S.~N.}\ \bibnamefont {Dorenbos}},\ and\ \bibinfo {author}
  {\bibfnamefont {V.}~\bibnamefont {Zwiller}},\ }\bibfield  {title} {\bibinfo
  {title} {Superconducting nanowire single-photon detectors: A perspective on
  evolution, state-of-the-art, future developments, and applications},\ }\href
  {https://doi.org/10.1063/5.0045990} {\bibfield  {journal} {\bibinfo
  {journal} {Appl. Phys. Lett.}\ }\textbf {\bibinfo {volume} {118}},\ \bibinfo
  {pages} {190502} (\bibinfo {year} {2021})}\BibitemShut {NoStop}%
\bibitem [{\citenamefont {Yue}\ \emph {et~al.}(2017)\citenamefont {Yue},
  \citenamefont {Gao}, \citenamefont {Lee}, \citenamefont {Kim},\ and\
  \citenamefont {Choi}}]{yue2017highly}%
  \BibitemOpen
  \bibfield  {author} {\bibinfo {author} {\bibfnamefont {W.}~\bibnamefont
  {Yue}}, \bibinfo {author} {\bibfnamefont {S.}~\bibnamefont {Gao}}, \bibinfo
  {author} {\bibfnamefont {S.-S.}\ \bibnamefont {Lee}}, \bibinfo {author}
  {\bibfnamefont {E.-S.}\ \bibnamefont {Kim}},\ and\ \bibinfo {author}
  {\bibfnamefont {D.-Y.}\ \bibnamefont {Choi}},\ }\bibfield  {title} {\bibinfo
  {title} {Highly reflective subtractive color filters capitalizing on a
  silicon metasurface integrated with nanostructured aluminum mirrors},\ }\href
  {https://doi.org/10.1002/lpor.201600285} {\bibfield  {journal} {\bibinfo
  {journal} {Laser Photonics Rev.}\ }\textbf {\bibinfo {volume} {11}},\
  \bibinfo {pages} {1600285} (\bibinfo {year} {2017})}\BibitemShut {NoStop}%
\bibitem [{\citenamefont {Bao}\ \emph {et~al.}(2020)\citenamefont {Bao},
  \citenamefont {Lin}, \citenamefont {Su}, \citenamefont {Zhou}, \citenamefont
  {Song}, \citenamefont {Li},\ and\ \citenamefont {Wang}}]{bao2020demand}%
  \BibitemOpen
  \bibfield  {author} {\bibinfo {author} {\bibfnamefont {Y.}~\bibnamefont
  {Bao}}, \bibinfo {author} {\bibfnamefont {Q.}~\bibnamefont {Lin}}, \bibinfo
  {author} {\bibfnamefont {R.}~\bibnamefont {Su}}, \bibinfo {author}
  {\bibfnamefont {Z.-K.}\ \bibnamefont {Zhou}}, \bibinfo {author}
  {\bibfnamefont {J.}~\bibnamefont {Song}}, \bibinfo {author} {\bibfnamefont
  {J.}~\bibnamefont {Li}},\ and\ \bibinfo {author} {\bibfnamefont {X.-H.}\
  \bibnamefont {Wang}},\ }\bibfield  {title} {\bibinfo {title} {On-demand
  spin-state manipulation of single-photon emission from quantum dot integrated
  with metasurface},\ }\href {https://doi.org/10.1126/sciadv.aba8761}
  {\bibfield  {journal} {\bibinfo  {journal} {Sci. Adv.}\ }\textbf {\bibinfo
  {volume} {6}},\ \bibinfo {pages} {eaba8761} (\bibinfo {year}
  {2020})}\BibitemShut {NoStop}%
\bibitem [{\citenamefont {Zhao}\ \emph {et~al.}(2019)\citenamefont {Zhao},
  \citenamefont {Duan}, \citenamefont {Li}, \citenamefont {Chen},\ and\
  \citenamefont {Zhang}}]{zhao2019integrating}%
  \BibitemOpen
  \bibfield  {author} {\bibinfo {author} {\bibfnamefont {X.}~\bibnamefont
  {Zhao}}, \bibinfo {author} {\bibfnamefont {G.}~\bibnamefont {Duan}}, \bibinfo
  {author} {\bibfnamefont {A.}~\bibnamefont {Li}}, \bibinfo {author}
  {\bibfnamefont {C.}~\bibnamefont {Chen}},\ and\ \bibinfo {author}
  {\bibfnamefont {X.}~\bibnamefont {Zhang}},\ }\bibfield  {title} {\bibinfo
  {title} {Integrating microsystems with metamaterials towards metadevices},\
  }\href {https://doi.org/10.1038/s41378-018-0042-1} {\bibfield  {journal}
  {\bibinfo  {journal} {Microsyst. Nanoeng.}\ }\textbf {\bibinfo {volume}
  {5}},\ \bibinfo {pages} {5} (\bibinfo {year} {2019})}\BibitemShut {NoStop}%
\bibitem [{\citenamefont {Padilla}\ and\ \citenamefont
  {Averitt}(2022)}]{padilla2022imaging}%
  \BibitemOpen
  \bibfield  {author} {\bibinfo {author} {\bibfnamefont {W.~J.}\ \bibnamefont
  {Padilla}}\ and\ \bibinfo {author} {\bibfnamefont {R.~D.}\ \bibnamefont
  {Averitt}},\ }\bibfield  {title} {\bibinfo {title} {Imaging with
  metamaterials},\ }\href {https://doi.org/10.1038/s42254-021-00394-3}
  {\bibfield  {journal} {\bibinfo  {journal} {Nat. Rev. Phys.}\ }\textbf
  {\bibinfo {volume} {4}},\ \bibinfo {pages} {85} (\bibinfo {year}
  {2022})}\BibitemShut {NoStop}%
\end{thebibliography}%

\end{document}